\documentclass[letterpaper,aps,prb,superscriptaddress,twocolumn]{revtex4}

\usepackage{graphicx}

\begin{document}

\title{The magnetic field influence on magnetostructural phase
transition in Ni$_{2.19}$Mn$_{0.81}$Ga}

\author{D.~A.~Filippov}

\affiliation{Physical Faculty, Moscow State University, Moscow
119899, Russia}

\author{V.~V.~Khovailo}

\affiliation{Institute of Fluid Science, Tohoku University, Sendai
980--8577, Japan}

\author{V.~V.~Koledov}

\affiliation{Institute of Radioengineering and Electronics,
Russian Academy of Sciences, Moscow 101999, Russia}

\author{E.~P.~Krasnoperov}

\affiliation{RRC, Kurchatov Institute, Moscow 123182, Russia}

\author{R.~Z.~Levitin}

\affiliation{Physical Faculty, Moscow State University, Moscow
119899, Russia}

\author{V.~G.~Shavrov}

\affiliation{Institute of Radioengineering and Electronics,
Russian Academy of Sciences, Moscow 101999, Russia}

\author{T.~Takagi}

\affiliation{Institute of Fluid Science, Tohoku University, Sendai
980--8577, Japan}

\begin{abstract}
Magnetic properties of a polycrystalline alloy
Ni$_{2.19}$Mn$_{0.81}$Ga, which undergoes a first-order
magnetostructural phase transition from cubic paramagnetic to
tetragonal ferromagnetic phase, are studied. Hysteretic behavior
of isothermal magnetization $M(H)$ has been observed in a
temperature interval of the magnetostructural transition in
magnetic fields from 20 to 100~kOe. Temperature dependencies of
magnetization $M$, measured in magnetic fields $H = 400$ and
60~kOe, indicate that the temperature of the magnetostructural
transition increases with increasing magnetic field.
\end{abstract}



\maketitle

Mn-containing Heusler alloys Ni$_{2+x}$Mn$_{1-x}$Ga have been
intensively studied during last years owing to their unique
combination of ferromagnetism and a structural (martensitic) phase
transition. Since martensitic transformation in Ni--Mn--Ga is of a
thermoelastic type, the materials exhibit a well-defined one- and
two-way shape memory effect~[1,2]. For the stoichiometric
Ni$_2$MnGa composition, martensitic transition temperature $T_m =
202$~K and Curie temperature $T_C = 376$~K~[3]. A phase diagram of
magnetic and structural transitions in Ni$_{2+x}$Mn$_{1-x}$Ga
alloys studied, theoretically and experimentally, in Refs.~[4,5].
It has been found, that a partial substitution of Mn for Ni leads
to a decrease of Curie temperature $T_C$ and an increase of the
martensitic transition temperature $T_m$ until they merge in a
compositional interval $x = 0.18 - 0.20$ (Fig.~1). The application
of an external magnetic field results in an increase of the
martensitic transition temperature [6--8]. If applied magnetic
field is strong enough, a reversible magnetostructural transition
can be realized in the magnetic field at constant temperature and
pressure~[9]. This means that the structural state of a sample can
be switched reversibly and that magnetic shape memory effect can
be realized by means of magnetic-field-induced martensitic
transition~[10,~11]. Although Ni$_{2+x}$Mn$_{1-x}$Ga alloys were
studied by a variety of methods, a systematic study of their
magnetic properties in the vicinity of the magnetostructural
transition is lacking. Magnetic properties of
Ni$_{2.18}$Mn$_{0.82}$Ga were studied in Ref.~[12], where
hysteretic behavior of temperature dependencies of magnetization
$M(T)$ has been observed. However, the anomalous field dependence
of magnetization $M$, reported in Ref.~[12], was found to be
sample dependent and, therefore needs further investigation. The
aim of this work is to investigate magnetic properties of
polycrystalline Ni$_{2.19}$Mn$_{0.81}$Ga alloy in strong magnetic
fields in the vicinity of the magnetostructural transition.

Polycrystalline Ni$_{2.19}$Mn$_{0.81}$Ga ingots were prepared by a
conventional arc-melting method in argon atmosphere. To attain a
good compositional homogeneity, the ingots were remelted three
times. Since weight loss during melting was approximately 0.1\%,
the composition of the ingots was assumed to be the nominal one.
The ingots were annealed in vacuum at 1050~K for 3~days. The
sample for measurements was cut from the middle part of the ingot.
Temperature and magnetic field dependencies of magnetization were
measured by a magnetometer equipped with a Hall sensor. Magnetic
fields up to 100~kOe were generated by a Bitter magnet.
Measurements of $M(H)$ were done in the following way. Initially
the sample was heated to $T = 380$~K, well above the temperature
of the transition. After that the sample was cooled down to a
desired temperature, which was controlled by a temperature
controller. In these measurements the magnetic field was changed
with a step of 10~kOe and was kept for some time in order to
insure that processes of structural transformation for a given
strength of the magnetic field had completely been finished.

\begin{figure}[h]
\begin{center}
\includegraphics[width=7cm]{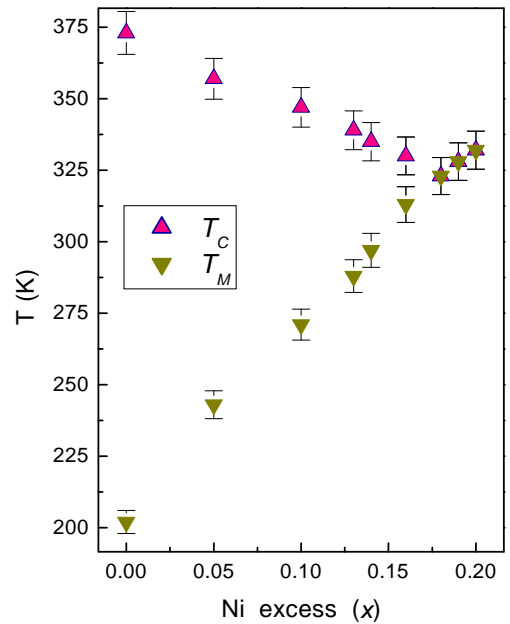}
\caption{The composition dependence of Curie temperature $T_C$ and
the martensitic transition temperature $T_m$ in
Ni$_{2+x}$Mn$_{1-x}$Ga ($x = 0 - 0.20$).}
\end{center}
\end{figure}

Typical temperature dependencies of magnetization $M$ in the
vicinity of the magnetostructural transition are shown in Fig.~2.
It is evident that as in the case of weak magnetic field $H =
400$~Oe (Fig.~2a), in a strong magnetic field $H = 60$~kOe
(Fig.~2b), the magnetization has a prominent temperature
hysteresis. Therefore, the magnetic transition in the
Ni$_{2.19}$Mn$_{0.81}$Ga alloy possesses characteristics typical
of a first-order phase transition. Under the action of the
magnetic field the characteristic temperatures of the
magnetostructural transition shift upward with a rate of $\approx
0.1$~K/kOe. This value agrees well with the data, published in
Refs.~[7,9,12].

\begin{figure}[t]
\begin{center}
\includegraphics[width=\columnwidth]{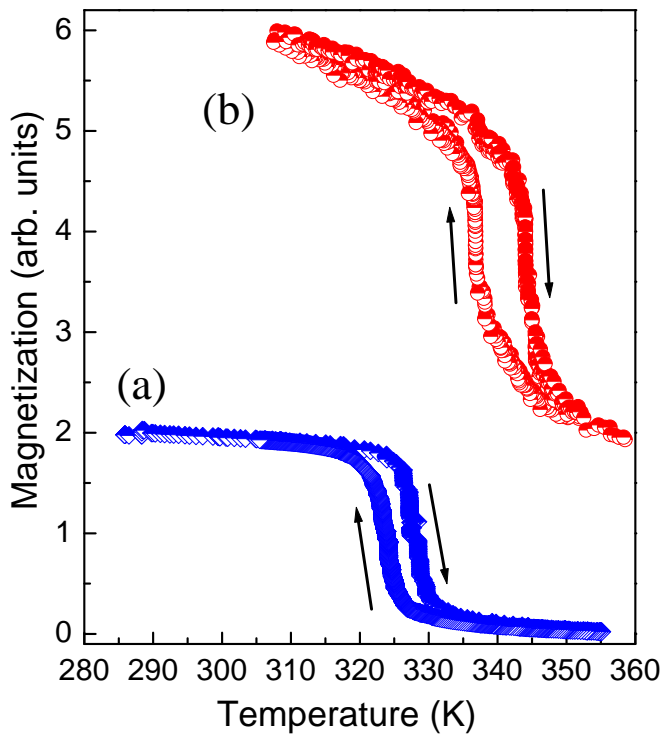}
\caption{Temperature dependencies of the magnetization of
Ni$_{2.19}$Mn$_{0.81}$Ga, measured in $H = 400$~Oe (a) and $H =
60$~kOe (b).}
\end{center}
\end{figure}

\begin{figure}[b]
\begin{center}
\includegraphics[width=\columnwidth]{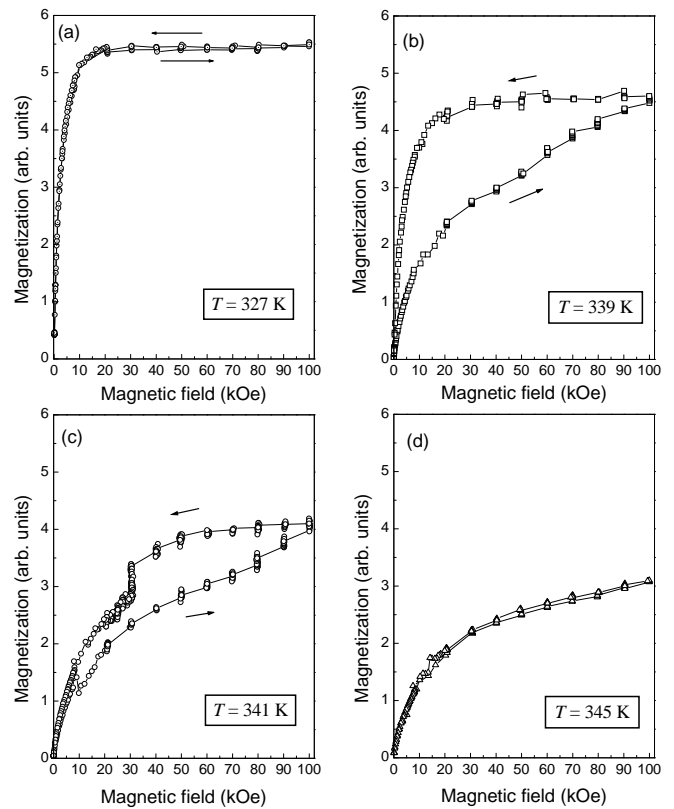}
\caption{Isothermal magnetization curves of
Ni$_{2.19}$Mn$_{0.81}$Ga measured in the vicinity of the
magnetostructural transition.}
\end{center}
\end{figure}

Isothermal magnetization curves taken at various temperatures in
the vicinity of the magnetostructural transition are shown in
Fig.~3. The $M(H)$ curves, measured at $T = 327$ (Fig.~3a) and
345~K (Fig.~3d) have field dependencies of magnetization typical
of a ferromagnet and a paramagnet, respectively. Contrary to this,
the $M(H)$ curves, measured in the temperature interval of the
mixed (martensite and austenite) state (Fig.~3b and 3c)
demonstrate hysteretic anomalies of magnetization in magnetic
field up to 100~kOe. As is seen from Fig.~3b, in the magnetic
field 0--20~kOe the behavior of the magnetization is similar to
the paramagnetic state. Further increase of the magnetic field
results in an anomalous growth of the magnetization, which
reaches, in a magnetic field of 100~kOe, 90\% of the magnetization
of the ferromagnetic state at $T = 327$~K (Fig.~3a). Upon
subsequent removal of the magnetic field, the magnetization
remains constant down to 20~kOe, where a drastic decrease, typical
for technical magnetization curve of a ferromagnet, is observed.
Based on the behavior of $M(H)$, measured at $T = 339$~K, it can
be suggested that induced by the magnetic field, low-temperature
(martensitic) ferromagnetic state of the sample remains stable
upon removal of the magnetic field.

The $M(H)$ dependencies, measured during application and removal
of magnetic field at $T = 341$~K are shown in Fig.~3c. At this
temperature the behavior of the magnetization is similar to the
paramagnetic state in magnetic fields up to 30~kOe. When the
magnetic field increases from 30 to 100~kOe, the magnetization
anomalously grows and reaches a value of 60\% of the magnetization
of the ferromagnetic state at $T = 327$~K. Upon subsequent
decrease of the magnetic field from 100 to 30~kOe, the attained
value of the magnetization is preserved. A rapid change of $M$
occurs at $H = 30$~kOe, where the magnetization drops to a value
of 110\% of the magnetization of the paramagnetic state at $T =
345$~K. During further decrease of the magnetic field from 30 to
0~kOe, the behavior of the magnetization is typical of the
paramagnetic state. The anomalous $M(H)$ dependencies measured at
$T = 341$~K can be explained as due to a reversible structural
transformation from paramagnetic austenite to ferromagnetic
martensite, induced by a magnetic field at a constant temperature.

The presented results of the magnetic measurements are in
accordance with experimental results, reported by Dikshtein
\textit{et al}.~[9], where a reversible magnetic-field-induced
structural phase transition was observed by an optical method. As
is seen from Fig.~3, induction of a structural transition by a
magnetic field is accompanied by unusual behavior of the
isothermal magnetization $M(H)$ in the temperature interval of the
coexistence of ferromagnetic martensite and paramagnetic
austenite. Actually, similar anomalies on $M(H)$ dependencies were
observed in Gd$_5$(Si$_x$Ge$_{4-x}$)~[13], which is another
representative of the intermetallic compounds with coupled
magnetostructural transition. Since such systems, with a close
relation between crystallographic structure and magnetism, are of
a great technological significance, further studies of
magnetostructural transitions in Ni$_{2+x}$Mn$_{1-x}$Ga~[12],
Co--Ni--Ga, and Co--Ni--Al~[14] alloys are of considerable
interest.

\bigskip

We are grateful to Professor A.~N.~Vasil'ev for helpful
discussions. This work was partially supported by the Grant of
RFBR--BRFBR 02--02--81030 Bel2002a, by the Grant of RFBR
02--02--16636a, and by the Grant-in-Aid for Scientific Research
(B) No.~11695038 from the Japan Society for the Promotion of
Science, and Izumi Science Technology Foundation.

\end{document}